\documentclass[journal]{IEEEtran}
\usepackage{graphicx}
\usepackage{array,multirow}
\usepackage{caption}
    \DeclareCaptionFont{mysize}{\fontsize{8}{8}\selectfont}
\usepackage{tabularx}
\usepackage{tabulary}
\usepackage{latexsym}
\usepackage{textgreek}
\usepackage{layout}
\usepackage{rotating}
\usepackage{amsmath,amsfonts,amssymb}
\usepackage{bm}
\usepackage{euscript}
\usepackage{algorithmic}
\usepackage{amssymb, amsmath}
\usepackage[numbers, square, comma, sort&compress]{natbib}
\usepackage[ruled,lined]{algorithm2e}
\usepackage{epstopdf}
\usepackage{gensymb}
\usepackage{subcaption}
\usepackage{siunitx}
\usepackage{url}
\usepackage{xcolor}
\usepackage{ragged2e}
\usepackage[english]{babel}
\usepackage{multicol}
\usepackage[official]{eurosym}
\usepackage{color}
\usepackage{acronym}
\usepackage{mathtools}
\usepackage[verbose]{wrapfig}
\usepackage{multirow}
\usepackage[normalem]{ulem}

\usepackage [autostyle, english = american]{csquotes}
\MakeOuterQuote{"}
\usepackage{caption} 
   \usepackage[belowskip=2pt, aboveskip=-13pt] {caption}
\DeclarePairedDelimiter{\ceil}{\lceil}{\rceil}

\newcommand{\seton}{\mathcal{N}_{\rm on}}
\newcommand{\setoff}{\mathcal{N}_{\rm off}}

\acrodef{API}[API]{Application Programmable Interface}
\acrodef{BS}[BS]{Base Station}
\acrodef{CNN}[CNN]{Convolutional Neural Network}
\acrodef{CPU}[CPU]{Central Processing Unit}
\acrodef{DETA-R}[DETA-R]{Dynamic and Energy-Traffic-Aware algorithm with Random behavior}
\acrodef{EB}[EB]{Energy Buffer}
\acrodef{EH}[EH]{Energy Harvesting}
\acrodef{ES}[ES]{Energy Saving}
\acrodef{EM}[EM]{Energy Manager}
\acrodef{ENAAM}[ENAAM]{Energy Aware and Adaptive Management}
\acrodef{E-LAN}[E-LAN]{Energy-Local Area Network}
\acrodef{GENM}[GENM]{Green-based Edge Network Management}
\acrodef{EPC}[EPC]{Evolved Packet Core}
\acrodef{ITS}[ITS] {Intelligent Transport System}
\acrodef{LOC}[LOC]{User Location Services} 
\acrodef{LLC}[LLC]{Limited Lookahead Control}
\acrodef{LS}[LS]{Location Service}
\acrodef{LSTM}[LSTM]{Long Short-Term Memory}
\acrodef{MEC}[MEC]{Multi-access Edge Computing}
\acrodef{ML}[ML]{Machine Learning}
\acrodef{MN}[MN]{Mobile Network}
\acrodef{TIM}[TIM]{Telecom Italia Mobile}

\acrodef{NOES}[NOES]{NO Energy Saving}
\acrodef{NFV}[NFV]{Network Function Virtualization}
\acrodef{NF}[NF]{Network Function}
\acrodef{PPG}[PPG]{Power Packet Grid}
\acrodef{QoS}[QoS]{Quality of Service}
\acrodef{RNN}[RNN]{Recurrent Neural Network}
\acrodef{RAN}[RAN]{Radio Access Network}
\acrodef{RMSE}[RMSE]{Root Mean Square Error}
\acrodef{RNN}[RNN]{Recurrent Neural Network}
\acrodef{TDM}[TDM]{Time Division Multiplexing}
\acrodef{UE}[UE]{User Equipment}
\acrodef{VLAN}[VLAN]{Virtual Local Area Network}
\acrodef{VM}[VM] {Virtual Machines}
\acrodef{VNF}[VNF]{Virtualized Network Function}

\begin{document}


\title{MEC-enabled Energy Cooperation for Sustainable 5G Networks Exploiting the Location Service API}

\author{\IEEEauthorblockN {Thembelihle Dlamini\IEEEauthorrefmark{1}}\\
	\IEEEauthorblockA {\IEEEauthorrefmark{1}Department of Electrical and Electronic Engineering \\ University of Eswatini, Manzini, Eswatini (Swaziland)\\}
	tldlamini@uniswa.sz \vspace{-0.4cm}
}

\maketitle
\thispagestyle{plain}
\pagestyle{plain}

\begin{abstract}

The substantial growth in wireless data traffic, and the emergence of \mbox{delay-sensitive} application/services requiring \mbox{ultra-low} latency, has resulted into a new Mobile Network (MN) design paradigm called {\it Multi-access Edge Computing (MEC)}. In this, the Base Stations (BSs) are empowered with computing capabilities, and they are densely deployed in order to increase network coverage and provide high throughput to mobile users. These developments require energy \mbox{self-sustainability} in order to minimize the carbon emission into the atmosphere and the dependence on the power grid. As a solution to this, we advocate for the integration of Energy Harvesting (EH) systems, e.g., solar panels or wind turbines (together with energy storage devices), into future BSs and edge computing systems (i.e., MEC servers). However, due to traffic load and harvested energy variations within a coverage area, the stored energy levels will also vary. To compensate for green energy imbalance within the network, {\it energy cooperation (transfer)} can be enabled by an energy trading application hosted in the MEC platform, and the energy packets traverse over the Power Packet Grid (DC power lines and switches) from the source BS(s) to the \mbox{energy-deficient} BS(s). In this paper, we jointly perform {\it energy allocation} and {\it energy routing} using an online algorithm based on Lyapunov \mbox{drift-and-penalty} optimization theorem (named Lyapunov) for enabling energy cooperation, leveraging the MEC Location Service (LS) Application Programmable Interface (API). Our numerical results reveal that the Lyapunov algorithm is able to deliver sufficient amount of energy under normal solar irradiance without the effects of the control parameter.

\end{abstract}


\IEEEpeerreviewmaketitle

\section{Introduction}

The use of distributed intelligence, whereby content, control, computation, are moved closer to mobile users (hereby referred to as the {\it network edge}), can help to improve network reliability and sustainability. This has lead to the emergence of \ac{MEC}, a new Mobile Network (MN) design paradigm that allows \acp{NF} to be virtualized and then deployed at the network edge in order to provide \mbox{ultra-low} latency services~\cite{etsimec_access}\cite{energymanagershow}. The network management can be achieved when the network intelligence is distributed deeper in the network, e.g., the MEC server can be placed at an aggregation point (a point in proximity to a group of \acp{BS} interconnected to the MEC server for computation offloading and BS system management). In addition, energy \mbox{self-sustainability} can be realized within the \ac{MN} through the use of \ac{EH} systems, e.g., solar panels and/or wind turbines, and energy storage systems. This result into an EH-powered MEC (EH-MEC) system and EH BSs. The presence of \ac{EH} systems and energy storage devices minimize the dependence on the power grid and the carbon emission into the atmosphere~\cite{energymanagershow}.

The EH BSs in proximity to the MEC server form an \ac{E-LAN}. Within the \mbox{E-LAN}, each BS is equipped with an energy storage device (\ac{EB}) for storing the harvested energy. In order to compensate for the imbalance in the harvested energy, caused by group mobility~\cite{mobility_dir} or traffic load variation, the surplus energy can be {\it transferred} from \mbox{BS-to-BS} through energy cooperation enabled by the MEC platform. Energy cooperation will be the key feature in future \acp{MN}. According to~\cite{sugiyama2012packet}, {\it energy transfer} can be accomplished using the \ac{PPG}, where power between sources and consumers is exchanged in the form of {\it packets} which flow from sources to consumers through power lines and electronic switches via the {\it energy router}~\cite{krylov2010toward} (which is responsible for the packets routing process). With the advent of \ac{NFV}, the energy router can be {\it softwarized (virtualized)} and then placed in the MEC server, in a form of an application, in order to enable \mbox{location-aware} energy routing. 
The user location context is provided by the \ac{LS} \ac{API}, which is a service that supports the mobile device location retrieval mechanism and then passes the information to authorized applications~\cite{etsimec_ls}, within the MEC platform.

Along the lines of energy cooperation, the following works have suggested different procedures. A matching \mbox{game-based} energy trading framework is presented in~\cite{matching_game}, where \acp{BS} with surplus energy are motivated to trade with other \acp{BS} with insufficient energy. The work of~\cite{Gambin2017} formulated optimal energy allocation and routing within a \ac{MN} as a convex optimization problem with the aim of improving the energy \mbox{self-sustainability} of the network, while achieving high energy transfer efficiencies under dynamic load and energy harvesting processes. An optimal assignment based on the Hungarian method is also presented. To mediate between the grid operator and a group of \acp{BS} to redistribute the energy packets, an aggregator is introduced in~\cite{xu2015}. Here, energy sharing between BSs is realized through the aggregator, i.e., one BS injects the surplus power while the other draws power from it. Lastly, in~\cite{Gurakan2013}, the authors investigate energy sharing control flows between BSs. To maximize the system throughput for all the considered network configurations, a directional \mbox{water-filling} based and an offline algorithm is considered.
It is worth observing that the aforementioned works are not considering the MEC platform as an enabler for energy cooperation. Thus, energy cooperation within the \ac{E-LAN} can be jointly achieved through energy allocation and routing procedures, leveraging the mobile device(s) location information provided by the \ac{LS} \ac{API}. 

\textbf{Objective and Contributions}: we consider the aforementioned scenario, where \ac{MEC} and \ac{EH} are combined into a single system located close to a group of EH BSs, towards energy self-sustainability in \acp{MN}. {\bf 1}) Motivated by the potential of EH and MEC, we introduce a paradigm shift in the \ac{PPG} network with the presence of the MEC server. The energy distribution is enabled by the energy cooperation and routing applications hosted in the virtualized MEC server. Then, {\bf 2}) we introduce the notion of {\it priority} \acp{EB}, i.e., the EBs of the BSs that are serving mobile users who are currently associated with the MEC server are maximized first. Lastly, {\bf 3}) we formulate a joint problem for {\it energy allocation} and {\it energy routing} as an online algorithm, taking into account the network imbalance caused by users mobility, i.e., group mobility, with the main goal of promoting the energy \mbox{self-sufficiency} of the BS system, which is realized using a virtualized router and an energy allocation algorithm leveraging the Lyapunov \mbox{drift-and-penalty} theorem. Here, energy cooperation decisions are made using only the currently available \ac{EB} levels and users' location information, obtained from the energy profiles/reports and the \ac{LS} \ac{API}, respectively.



   
   
   

The proposed optimization strategy leads to a considerable energy transfer under the guidance of the energy router application, promoting self-sustainability with the mobile network through the use of green energy.

The rest of the paper is structured as follows: Section~\ref{sec:sys} discuss the system model and Section~\ref{sec:prob} provide the optimization problem and the online algorithm. Section~\ref{sec:perf} evaluate our online energy cooperation procedures, while Section~\ref{sec:concl} gives concluding remarks. 

\section{System Model}
\label{sec:sys}

\begin{figure}[t]
	\centering
	\includegraphics[width = 0.45\textwidth]{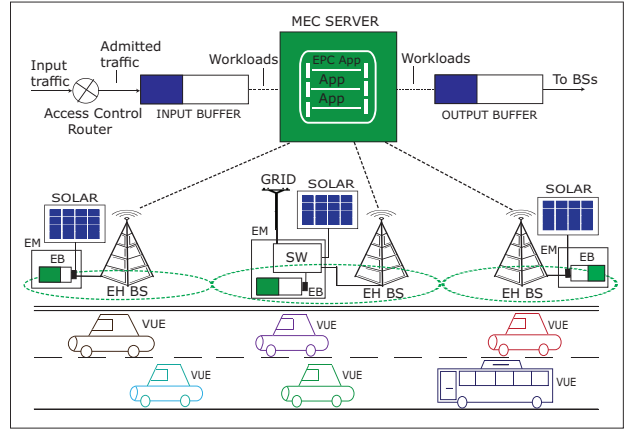}
	\caption{Edge network topology. The electromechanical switch (SW) is responsible for selecting the appropriate source of energy.}
	\label{fig:edge_computing}
\end{figure}

As a major deployment of MEC~\cite{etsimec_access}, the considered network scenario is illustrated in Fig.~\ref{fig:edge_computing}. It consists of a \mbox{densely-deployed} \ac{MN} featuring $N = |\mathcal{N}|$ \acp{BS} ($\seton$ for \mbox{on-grid} set and $\setoff$ for \mbox{off-grid} set), and a \mbox{cache-enabled}, TCP/IP \mbox{offload-enabled} ({\it partial} computation at the network adapter), virtualized MEC server. The MEC server is assumed to be deployed at an aggregation point~\cite{etsimec_access}\cite{energymanagershow} for purposes of energy cooperation, computation offloading, resource centralization and BS system management without a significant amount of latency. Also, it is assumed to be equipped with higher computational and storage resources compared to the \mbox{end-user} device. The server clients are assumed to be mobile users moving in groups, hereby referred as Virtual User Equipment (VUE), and they are represented by the Reference Point Group Mobility Model (RPGM~\cite{mobility_dir}). Their current locations are known through the \ac{LS} \ac{API}~\cite{etsimec_ls}, within the MEC platform.
Each site, i.e., the \ac{BS} or MEC platform, is empowered with \ac{EH} capabilities through a solar panel and wind turbines (not shown in Fig.~\ref{fig:edge_computing}), and an \ac{EB} that enables energy storing. Energy supply from the power grid is also available to some BSs. Moreover, the \ac{EM} is an entity responsible for selecting the appropriate energy source and for monitoring the energy level of the \ac{EB}.
Similar to our previous work~\cite{comp_plus_comm_mec}, the router is virtualized and locally hosted as an application. It is responsible for energy routing processes. The \ac{MN} is overlaid on top of the \ac{PPG} network.
In addition, we consider a \mbox{discrete-time} model, whereby time is discretized as \mbox{$t = 1,2,\dots$}, and each time slot $t$ has a fixed duration $\tau$.

\subsection{Power Packet Grid}
\label{sub:ppg}

Power packets distribution is enabled by the use of the Power Packet Grid (PPG) infrastructure, where the MEC server is in charge of making energy routing and power allocation decisions. The packetized power transmits in a \ac{TDM} manner over direct current (DC) power links (electric wires) that connect BSs~\cite{ma2017optimal}. In TDM systems, the power channel becomes a limiting factor thus each power link can only be used for a single energy trading operation at a time. The PPG network incurs power losses due to the resistance of the considered transmission medium between source and destination. Here, we use the following model for the resistance~\cite{von2006electric}: $ {\rm \Upsilon} = \frac{\rho \ell}{\rm A}$, where $\rho$ is the resistivity of the wire in $\SI{}{\Omega \milli\milli^2 / \meter}$, $\ell$ is the length of the power link in meters, and $\rm A$ is the \mbox{cross-sectional} area of the cable in $\SI{}{\milli\milli^2}$. Finally, we assume that all power links are of the same type, with the same $\rm A$. The presence of \mbox{high-performance} switches and routers within the PPG network helps to speed power packet processing.

\subsection{Communication traffic and Energy consumption}
\label{sub:serverload}

\begin{figure}[t]
	\centering
	\resizebox{\columnwidth}{!}{\input{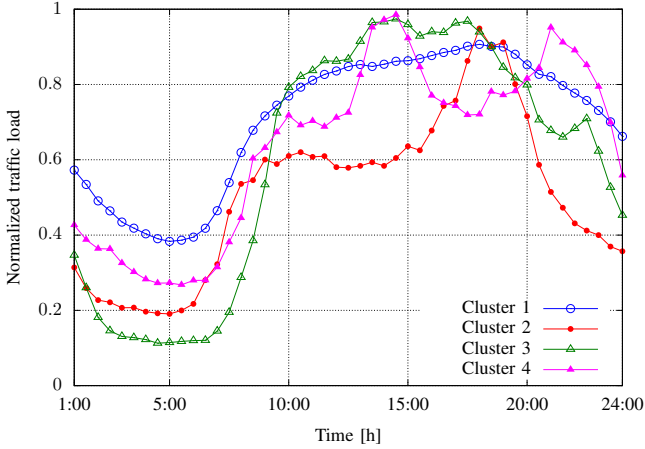}}
	\caption{Example traces for normalized BS traffic loads. The data from~\cite{bigdata2015tim} has been split into four representative clusters.}
	\label{fig:trace_load}
\end{figure}

Traffic volume at individual \acp{BS} can be estimated using historical mobile traffic traces. In this paper, real \ac{MN} traffic load traces obtained from the Big Data Challenge organized by \ac{TIM}~\cite{bigdata2015tim} are used to emulate the traffic load.
Specifically, the used data was collected in the city of Milan during the month of November 2013, and it is the result of user interaction within the \ac{TIM} MN, based on Call Detail Record (CDR) files. Each CDR file consists of SMS, Calls and Internet records.
To understand the behavior of the mobile data, the clustering algorithm \mbox{X-means}~\cite{pelleg2000x} has been applied to classify the load profiles into several categories. In our numerical results, each \ac{BS} $n$ has an associated load profile ${L}_{n}(t)$, which is picked at random as one of the four clusters in Fig.~\ref{fig:trace_load}. Moreover, we assume that ${L}_{n}(t)$ consists of $80\%$ computation workloads and the remainder is standard workloads (i.e., \mbox{delay-tolerant} traffic).

The total energy consumption ([$\SI{} {\joule}$]) from BS $n$ at time slot $t$ is formulated as follows, inspired by~\cite{mec_lyapunov}\cite{online_pimrc}:
\begin{equation}
	\mbox{$\theta_{{\rm BS},n}(t) = \theta_{0} + \theta_{{\rm load},n}(t)$}\,,
	\label{eq:bsconsupt}
\end{equation}
where $\theta_{0}$ is a constant value (\mbox{load-independent}), representing the operation energy which includes baseband processing, radio frequency power expenditures, etc. $\theta_{{\rm load},n}(t)$ represents the total wireless transmission (load dependent) power to meet the target transmission rate from the BS to the served user(s) and to guarantee low latency at the edge. Since we assume a \mbox{noise-limited} channel and the guarantee of low latency requirements at the edge, $\theta_{{\rm load},n}(t)$ is obtained by using the transmission model in~\cite{mec_lyapunov} (see Eq.~(5) in this reference). 
%
%
We remark that the MEC server energy consumption is not considered in this work as its role is to enable energy cooperation only, we refer the reader to our previous work in~\cite{comp_plus_comm_mec} for details about MEC server energy consumption.

\subsection{Energy Patterns and Storage}
\label{sub:eebuffer}

\begin{figure}[t]
	\centering
	\resizebox{\columnwidth}{!}{\input{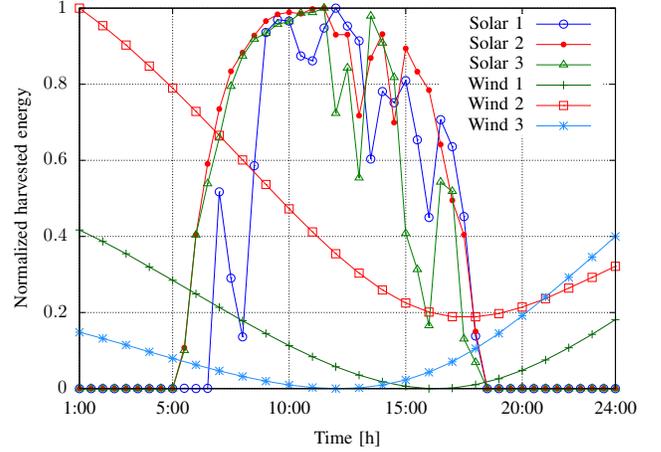}}
	\caption{Example traces for harvested solar traces from~\cite{amerinia} and wind traces from~\cite{belgium}.}
	\label{fig:energy_trace}
\end{figure}

The energy buffer is characterized by its maximum energy storage capacity $\beta_{\rm max}$, and power charging/discharging and leaking losses are not assumed. At each time slot $t$, the \acp{EM}, from BS sites, provide the energy level reports to the MEC server through the pull mode procedure (e.g., File Transfer Protocol (FTP)), thus the \ac{EB} level $\beta(t)$ is known, enabling the energy cooperation process. 
In this work, the amount of harvested energy $H_n(t)$, per site (BS $n$), in time slot $t$ is obtained from \mbox{open-source} solar traces within a solar panel farm located in Armenia~\cite{amerinia} and also wind traces within a wind farm located in Belgium~\cite{belgium} ({\it see} Fig.~\ref{fig:energy_trace}). The data in the dataset is aggregated to match our time slot duration ($\SI{1} {\minute}$). The dataset is the result of daily environmental records for a place assumed to be free from surrounding obstructions (e.g., buildings, shades). In our numerical results, $H_n(t)$ is obtained by picking a day at random in the dataset and then scaling the solar energy to fit the \ac{EB} capacity $\beta_{\rm max}$ of $\SI{490} {\kilo\joule}$. Here, wind energy is selected as a source during the solar energy \mbox{off-peak} periods. 
The available \ac{EB} level $\beta_{n}(t + 1)$  for an \mbox{off-grid} BS $n \in \setoff$ in time slot $t+1$ is calculated as follows: 
\begin{equation}
\beta_{n}(t + 1) = \min\{\beta_{n}(t) + H_{n}(t) - \theta_{{\rm BS},n}(t) + G_{n}(t),\beta_{\rm max}\},
\label{eq:offgrid}
\end{equation}
\noindent where $\beta_{n}(t)$ is the energy level in the battery at the beginning of time slot $t$, $\theta_{{\rm BS},n}(t)$ is the energy consumption of the communication site over time slot $t$, see Eq.~(\ref{eq:bsconsupt}), and $G_{n}(t)$ is the amount of energy transferred during time slot $t$, which can either be positive (BS $n$ is a consumer) or negative (BS $n$ is a source).
%
The energy level  of an \mbox{on-grid} BS $n \in \seton$ is updated as:
\begin{equation}
	\beta_{n}(t + 1) = \min\{\beta_{n}(t) + H_{n}(t) -  \theta_{{\rm BS},n}(t) + G_{n}(t) + E_{n}(t), \beta_{\rm max}\},
	\label{eq:ongrid}
\end{equation}
where the new term $E_{n}(t) \geq 0$ represents the energy purchased by BS $n$ from the power grid during time slot $t$. We remark that $\beta_n(t)$ is updated at the beginning of time slot $t$ whereas $H_n(t)$, $\theta_{{\rm BS},n}(t)$ and is  only known at the end of it. In addition, \mbox{on-grid} BSs are always energy sources.

For decision making in the MEC server, the received \ac{EB} level reports are compared with the following thresholds: $\beta_{\rm low}$ and $\beta_{\rm up}$, respectively termed the lower and upper energy threshold with $0 < \beta_{\rm low} < \beta_{\rm up} < \beta_{\rm max}$, and from the comparison the behavior of a BS is determined. $\beta_{\rm up}$ corresponds to the desired \ac{EB} level and $\beta_{\rm low}$ is the lowest energy level that any \ac{BS} should ever reach. At time slot $t$, if $\beta_{n}(t) > \beta_{\rm up} $, then BS $n$ behaves as an {\it energy source} and it is eligible for trading an amount of $\beta_{n}(t) - \beta_{\rm up} $ to other BSs. If $\beta_{n}(t) < \beta_{\rm low}$, then BS $n$ becomes an {\it energy consumer}, its energy demand amounts to $d_{n}(t) = \beta_{\rm low} - \beta_{n}(t)$ so that its energy buffer would ideally become equal to the lower threshold $\beta_{\rm low}$ by the end of the current time slot. Moreover, if the total energy in the buffer at the end of the current time slot, $t$, is $\beta_{n}(t) < \beta_{\rm up}$ and the BS $n$ is \mbox{on-grid}, then the difference $E_{n}(t) = \beta_{\rm up} - \beta_{n}(t)$ is purchased from the power grid in slot $t$. 
%
Our optimization framework in Section~\ref{sec:prob} makes sure that $\beta_n(t)$, never falls below $\beta_{\rm low}$ and guarantees that \mbox{$\beta_{\rm up}$} is reached at every time slot. 

\section{Problem Formulation} 
\label{sec:prob}

In this section, we formulate an optimization problem to obtain {\it energy cooperation} through {\it energy allocation} and {\it routing} within an \mbox{E-LAN}, taking into account mobile device(s) location information and the association with the MEC server. 

\subsection{Notations}
\label{sub:not}

We use the indices $s$ and $c$ to denote an arbitrary energy source and energy consumer. Then, $\mathcal{L}_{s} = \{1,\dots,s,\dots, S\}$ and $\mathcal{L}_{c} = \{1,\dots,c,\dots, C\}$, represents a set of BSs acting as sources and consumers, respectively. With $v_{s,c}$ we mean the total amount of energy available to transfer from source $s$ to consumer $c$. This depends on the distance between $s$ and $c$, and the associated distribution losses. In matrix notation we have ${\bm V}$  = [$v_{s,c}$]. With $x_{s,c} \in [0,1]$, we mean the fraction of $v_{s,c}$ that is actually delivered at the consumer $c$ from source $s \in \mathcal{L}_{s} $. In matrix notation $\bm X$ = [$x_{s,c}$]. $d_{n}(t)$ represents the energy demand of BS $n$, and $\mu_{s,c}$ represents the number of hops in the energy routing topology between source $s \in \mathcal{L}_{s} $ and consumer $c \in \mathcal{L}_{c} $. $j_{s,c}(t)$ refers to the duration at which the power link is occupied, named {\it link occupancy duration}, and $\Delta$ is the MEC server's response time, i.e., the maximum time allowed for the computation and communication processes. 

\subsection{Optimization Problem}
\label{sub:obj_functions}

Our goal is to maximize the \ac{EB} levels of the \acp{BS}. Here, we introduce the notion of {\it priority} \acp{EB}, i.e., the EBs of the BSs that are serving UEs who are currently associated with the MEC server are maximized first. We assume each UE is served by one BS at a time. In this way, an optimal energy allocation is performed by first identifying \acp{BS} where the buffered service requests are from and to, followed by determining the suitable energy sources with minimum $\mu_{s,c}$ for each target BS, denoted by $\beta_{{\rm obj},n}(t+1)$, and then routing the energy packets from $s$ to $c$ using the minimum number of \mbox{mini-slots}, i.e., reducing the link occupancy $j_{s,c}(t)$. In this work, to map the BSs and the buffered workloads we group all the BSs that are currently associated with the mobile edge host, and they are denoted by $I \subset \mathcal{N}$.
Then, our objective function is defined as follows:\footnote{For notation simplicity, we retain the $\beta$ notation even in our objective as $\beta_{{\rm obj},n}(t+1)$, since we maximize the priority EBs first.}
\begin{eqnarray}
        \label{eq:objt}
        \textbf{P1} & : & \max_{\mathcal{E}} \sum_{n = 1}^{N}\sum _{t=1}^{T} \beta_{{\rm obj},n}(t+1) \\
        && \hspace{-1.25cm}\mbox{subject to:} \nonumber \\
        {\rm C1} &:  & j_{s,c} (t) \leq \Delta, \nonumber \\
        {\rm C2} & : & \beta_{n}(t) \geq \beta_{\rm low}, \quad t=1,\dots, T, \, \nonumber
\end{eqnarray}
where $\mathcal{E} = \{j_{s,c}, \mathcal{L}_{c}, \mathcal{L}_{s}\}$ is the set of objective variables to be configured at slot $t$ by the energy trading application (in the MEC server), for the energy cooperation processes. The constraints C1 guarantees the \mbox{real-time} performance in the energy routing process and C2 makes sure that the \ac{EB} level is always above or equal to a preset threshold $\beta_{\rm low}$, to guarantee {\it energy \mbox{self-sustainability}} over time. 
To solve $\rm {\bm P1}$ in Eq.~(\ref{eq:objt}), we make use of the Lyapunov \mbox{drift-plus-penalty} theorem~\cite{lyapunov} and heuristics. 

\subsection{Online Energy Cooperation}
\label{sub:online}

In this subsection, an online algorithm is presented to solve {\rm P1}. 
In subsection~\ref{energycoop}, we solve {\rm P1} by first constructing an energy deficit queue, then expressing the energy queue as a Lyapunov \mbox{drift-plus-penalty} equation. We then perform the energy allocation and routing process, proposing two heuristics. 



\subsubsection{Energy Cooperation online algorithm}
\label{energycoop}

Our algorithm, called algorithm~\ref{lyapunov}, solves \textbf{P1} based on Lyapunov optimization~\cite{lyapunov}. The algorithm is purely online and requires only currently available information as inputs, i.e., $I, \mathcal{L}_{s}, \mathcal{L}_{c}$. Specifically, we introduce the EB queue $\beta_{{\rm obj},n}(t)$, i.e., the current \ac{EB} level of BS $n$, and then we assume that the initial state \mbox{$\beta_{{\rm obj},n}(0) = 0, \forall \, n \, \in \mathcal{N}$}. The EB queue evolves according to the following queuing dynamics equation:
\begin{equation}
		\beta_{{\rm obj},n}(t+1) = \max [\beta_{{\rm obj},n}(t) + V(t) - \beta_{\rm max}, 0], \,  
		\label{eq:buffer_max}
\end{equation}
where $V(t)$ is the energy packets delivered at a fractional cost $x_{s,c}$ (i.e., it represents the maximum number of power packets that can be reliably transferred over the power channel destined for BS $n$), and $\beta_{\rm max}$ is the upper \ac{EB} level bound that cap the buffer energy (here, by intuition \mbox{$\beta_{{\rm obj},n}(t) + V(t) \leq \beta_{\rm max}$}). 

To solve Eq.~(\ref{eq:buffer_max}), we express the equation as Lyapunov \mbox{drift-plus-penalty} equation (see \textbf{P2} in Algorithm~\ref{lyapunov}) where $\beta_{{\rm obj},n}(t) \, \theta_{{\rm BS},n}(t)$ represents the energy drift and $\Lambda \, V(t)$ represents the energy transfer penalty cost, with $\Lambda \geq 0$ as a fixed penalty control parameter (sometimes referred to as the important weight). $\Lambda$ makes a dynamical \mbox{trade-off} between energy transfer penalty minimization and energy drift, thus minimizing the weighted penalty term through an energy allocation policy of choosing the source $s$ with minimum $\mu_{s,c}$. Minimizing the Lyapunov drift has a goal of pushing queues to a lower congestion state, hence making the network stable, while for the penalty function the network evolve towards optimal values. The value of $\Lambda$ depends on specific modeling parameters and it is determined on \mbox{trial-and-error} basis, as it cannot be determined in advance. Theorem $1$ provides the performance guarantee of algorithm~\ref{lyapunov}. \\

\textbf{Theorem 1.} By applying Algorithm~\ref{lyapunov}, the time averages of the penalty process satisfies:
\begin{equation}
		\lim_{T\to\infty} \frac{1}{T} \sum_{t=1}^{T}(\mathbb{E}\{V(t)\} \leq V^{*}(t) + \frac{1}{2\Lambda} (\sum_{n=1}^{N} \bar{\beta}_{{\rm obj},n}(t) - \beta_{\rm max})^{2})\,,
		\label{eq:theorem}
\end{equation}
as we desire to make $V(t)$ be less than or equal to some target value $V^{*}(t)$ and $\bar{\beta}_{{\rm obj},n}(t)$ is the average of the \ac{EB} levels along the VUEs trajectory. For the sake of brevity, we left out the {\it Proof} of this theorem (for more details see~\cite{mec_lyapunov}).
In theorem $1$, the time averages of the penalty is upper bounded by the target value $V^{*}(t)$ plus a constant. The constant depends on the control parameter $\Lambda$, which makes a trade-off between penalty minimization and energy drift.\\

Next, we discuss the energy allocation and routing heuristics algorithms, as follows:\\

\subsubsection*{\bf Energy allocation} this process takes place after observing a \acp{BS} associated with the MEC server at $t$, i.e., set $I$. The algorithm (named Algorithm~\ref{lyapunov}) to find the closest source proceeds as follows (beginning from line 03 algorithm~\ref{lyapunov}). For each $\beta_{{\rm obj},n} \in I$, (i) check if $ \beta_{{\rm obj},n} \in \mathcal{L}_{c}$ and $ \beta_{{\rm obj},n} \in \setoff$. If \mbox{off-grid} and energy deficient, compute the energy demand $d_{n}(t)$, else do nothing ($\beta_{{\rm obj},n}$ is \mbox{on-grid}). From the received energy reports from the EM, the $v_{s,c}$ values for source $s \in \mathcal{L}_{s}$, and their $\mu_{s,c}$ values, are determined w.r.t consumer $c$. The sources are then classified into two sets, one for sources with $v_{s,c} \geq d_{n}(t)$ and the other with $v_{s,c} < d_{n}(t)$. Each set is sorted in ascending order, the source with minimum $\mu_{s,c}$ first. From the set with $v_{s,c} \geq d_{n}(t)$, the energy source candidate with minimum $\mu_{s,c}$ (in terms of hops only) is selected to transfer a certain amount of $v_{s,c} = (\beta_{n}(t) - \beta_{\rm low}) \cdot \mu_{s,c}$ ([$\SI{} {\joule}$]) to consumer $c$. If the set with $v_{s,c} \geq d_{n}(t)$ is empty, the source is selected from the set with $v_{s,c} < d_{n}(t)$. This may give rise to {\it connection outages} due to an insufficient amount of energy being transfer (the solution has been left for our future work). Then, the control parameter $\Lambda$ (fixed value) is applied to the drift-plus-penalty equation (see \textbf{P2} in algorithm~\ref{lyapunov}), followed by an update  of Eq.~(\ref{eq:buffer_max}). (ii) Lastly, return the selected source $s$ and the allocated $v_{s,c}$, as  $\bm X$.\\

\begin{small}
\begin{algorithm}[t]
\begin{tabular}{l l}
{\bf Input:} & $I, \Delta, \mathcal{L}_{s}, \mathcal{L}_{c}$\\
{\bf Output:} & all sources $\bm X$ with minimum $\mu_{s,c}$ and $v_{s,c}$\\
01:		&\hspace{-1cm} $\beta_{{\rm obj},n}(0) \leftarrow 0$\\
02: 	&\hspace{-1cm} {\bf For} $t = 1$ to $T$ {\bf do} \\
    	&\hspace{-1cm} \quad Observe $I$ at the beginning of each $t$\\
03: 	&\hspace{-1cm} \quad {\bf For} each $\beta_{{\rm obj},n} \in I$ {\bf do}\\
		& \hspace{-0.5cm} \quad {\bf If} $ \beta_{{\rm obj},n} \in \mathcal{L}_{c}$ and $ \beta_{{\rm obj},n} \in \setoff$ {\bf then}\\
    	&\hspace{-0.5cm} \qquad Compute $d_{n}(t)$\\
    	&\hspace{-0.5cm} \qquad ${\bm V} \leftarrow$ choose source(s) $s \in \mathcal{L}_{s}$ with \\    
		&\hspace{-0.1cm} \qquad minimun $\mu_{s,c}$, in order to minimize\\
		&\hspace{-0.1cm} \qquad $\textbf{P2}: \beta_{{\rm obj},n}(t)\, \theta_{{\rm BS},n}(t) + \Lambda \, V(t) $\\
    	&\hspace{-0.5cm} \qquad Update $\beta_{{\rm obj},n}(t+1)$\\
04:		&\hspace{-1cm} \quad{\bf End}\\
05:		&\hspace{-1cm} {\bf End}\\
06: 	&\hspace{-1cm} {\bf Return} $\bm X \leftarrow $ return sources\\
\end{tabular}
\caption{Lyapunov-based Energy Allocation}
\label{lyapunov}
\end{algorithm}
\end{small}

\subsubsection*{\bf Energy routing} the routing process performed in the MEC server ensures the delivery of $x_{s,c}v_{s,c}$ ([$\SI{} {\joule}$]) from source $s$ to consumer $c$ (i.e., $\beta_{{\rm obj},n}$), over the PPG network. It is worth noting that the routing algorithm is executed at the beginning of each time slot, when a new allocation matrix $\bm X$ is returned by Algorithm. From a static routing table, the route path from source $s$ to consumer $c$ is obtained, as $h_{s,c}$, for each target BS. The power packets are routed from source $s$ to $\beta_{{\rm obj},n}$ via the MEC server, i.e., the routing application. For each energy transfer operation, a single link is used, and power packets are transmitted in \mbox{mini-slots}. In this paper, we want to minimize the time it takes to deliver $x_{s,c}v_{s,c}$ at the consumer $c$, that is, the duration of occupying the single link, $j_{s,c}(t)$, in order to guarantee the \mbox{real-time} performance expected in MEC server. Having minimum $j_{s,c}(t)$ will assist in handling possible sudden location changes of the VUEs. For a given maximum energy transmission capacity $\varphi_{\rm max}$ for a power link, the number of \mbox{mini-slots} required for the transfer of $x_{s,c}v_{s,c}$, is obtained similar to~\cite{Gambin2017} as $y_{s,c} = \ceil[\big]{(x_{s,c}v_{s,c})/ \varphi_{\rm max}}$. From this, we can observe that $j_{s,c}(t) = y_{s,c} + \xi$, where $\xi$ represents the processing time and buffering delays in the MEC server that we are expecting in a queuing system (assumed as $\SI{2} {\second}$ in our case). 

The routing algorithm proceeds as follows: (i) For each $\beta_{{\rm obj},n}\in I$ and $\beta_{{\rm obj},n} \in \setoff$, a route $h_{s,c}$ is obtained from the static routing table for the transmission of the power  packets from source $s$ to consumer $c$, (ii) then $y_{s,c}$ is determined, (iii) the link occupancy timer $j_{s,c}(t)$ is initialized, (iv) packet transmission takes place (the energy transfer is achieved using route $h_{s,c}$ for a number of \mbox{mini-slots} $y_{s,c}$) after sending the $\beta_{{obj},n}$ ip address (in our case we use the BS id = $n$) to the source $s$, (v) after each buffer deque check if $j_{s,c}(t) \leq \Delta$, (vi) If the timer is still within bounds, continue packet transmission process until completion, while monitoring $j_{s,c}(t)$, else continue to deliver all the packets (in this way we avoid outages at a $j_{s,c}(t)$ cost), (vii) After power packet transmission completion, release the power link for reuse. 

\section{Performance Evaluation}
\label{sec:perf}

In this section, we show some selected numerical results for the scenario of Section~\ref{sec:sys}. The parameters that were used for the simulations are listed in Table~\ref{tab_opt}.

\begin{table}[t]
\caption{System Parameters.}
\center
\begin{tabular} {|l| l|l|}
\hline 
{\bf Parameter} & {\bf Value} \\ 
\hline
Number of BSs, $\mathcal{N}$ & $24$ (5 $\seton$)\\
Cable resistivity, $\rho$ &  $\SI{0.023}{\Omega \milli\milli^2/\meter}$\\
Cable cross-section, ${\rm A}$ &  $\SI{10}{\milli\milli^2}$\\
Length of a power link, $\ell$ &  $\SI{100} {\meter}$\\
Energy storage capacity, $\beta_{\rm max}$ & $\SI{490} {\kilo\joule}$\\
Higher energy threshold, $\beta_{\rm up}$  & $70$~\%\\
Lower energy threshold, $\beta_{\rm low}$  & $30$~\%\\
Max. server processing time, $\Delta$ & $\SI{60} {\second}$ \\
mini slot duration & $\SI{5} {\second}$\\
Transmission capacity, $\varphi_{\rm max}$ & $\SI{100} {\kilo\joule}$/mini slot\\
\hline 
\end{tabular}
\label{tab_opt}
\end{table}

\textbf{Simulation Setup}: we consider a $4$ x $6$ grid of square regions covered by $24$ \mbox{densely-deployed} \acp{BS} located on the grid intersection, with overlapping coverage areas. \mbox{On-grid} BSs \mbox{co-exist} with \mbox{off-grid} BSs. Here a \mbox{two-lane} highway vehicular environment is simulated, where $10$ vehicles (VUEs), each having UEs onboard, proceed without making abrupt \mbox{U-turns}. Our time slot is set to $\tau = \SI{1} {\minute}$. For our simulation, we use {\it Python} as the programming language.

\begin{figure}
    \centering
	\includegraphics[width = \columnwidth]{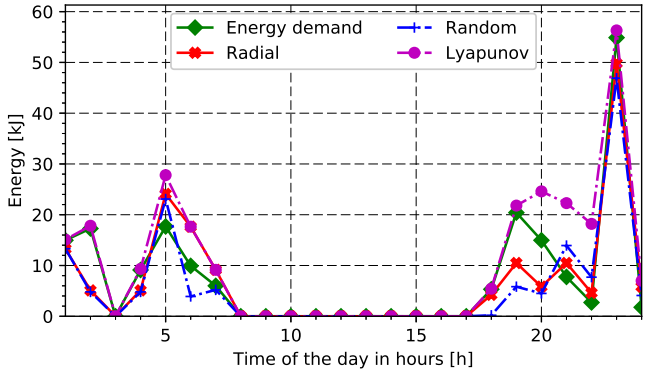}
	\caption{Average transferred energy along the VUE trajectory for one day observation under same solar irradiance, with $\Lambda = 1$. }
	\label{fig:control_trans}
\end{figure}

\textbf{Numerical results}: Fig.~\ref{fig:control_trans} shows the average energy transferred and energy demand $d_{n}(t)$ (named {\bf Energy demand}, see green curve) over time, under normal solar irradiance. We benchmark our proposed energy cooperation strategy (named {\bf Lyapunov}) with the following two benchmarks that employs a different way for obtaining the energy source $s$: \textbf{i)} A radial source search approach (named {\bf Radial}): a radial search is performed following~\cite{radial}. We first check if the four neighbors of the target \ac{BS} are in $\mathcal{L}_{s}$. Otherwise, we check if the \mbox{neighbors-of-neighbors} of the target BS are sources. Our search is limited to only two iterations. Then, \textbf{ii)} A random source search approach (named {\bf Random}): the source is randomly selected from $\mathcal{L}_{s}$.

On average, the Lyapunov algorithm is able to meet the energy demand and transfer an energy surplus of $38$~\% (\mbox{$\SI{5} {\hour} - \SI{7} {\hour}$}), and an energy surplus of $39$~\% (\mbox{$\SI{18} {\hour} - \SI{24} {\hour}$}). Thus, the proposed algorithm, Lyapunov, is able to improve the \ac{EB} level at each time instance if $\beta_{{\rm obj},n}$ is energy deficient, when compared with the other algorithms. Considering the highest energy demand peak ($\SI{23} {\hour}$), the energy delivered by the Lyapunov algorithm is $12$~\% higher than the energy delivered by the Radial algorithm and $17$~\% higher than the Random algorithm. Both benchmarks, Radial and Random, transferred energy below demand $d_{n}(t)$ and this poor performance is due to the limited source exploration.

\begin{figure} [t]
	\centering
	\includegraphics[width = \columnwidth]{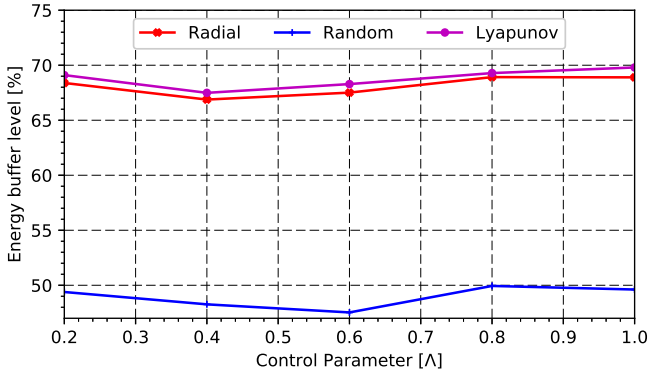}
	\caption{Control parameter $\Lambda$ impact towards \ac{EB} level update under normal solar irradiance.}
	\label{fig:controllambda}
\end{figure}

The impact of the control parameter $\Lambda$ is shown in Fig.~\ref{fig:controllambda}. From the figure, we can observe the variation of the average \ac{EB} level percentage in response to the value of $\Lambda = 0.2, 0.4, 0.6, 0.8, 1$ (satisfying $\Lambda > 0$). Results show that $\Lambda$ introduces a lower influence on the weighted penalty term in the control decision and allows a smooth tradeoff between the energy drift and penalty minimization. From this figure, it can confirmed again that the average \ac{EB} level with Lyapunov is higher than the one applying the two benchmarks.

The achieved energy cooperation performance results has been achieved under the guarantee of link occupancy duration and the usage of minimal number of \mbox{mini-slots}.

\section{Conclusion}
\label{sec:concl}

In this paper, we have envisioned  an environment where densified small cells base stations are capable of energy harvesting and performing energy cooperation processes, enabled by the MEC server placed at an aggregation point, whereby the power packets traverse over the Power Packet Grid. The combination of energy harvesting, energy cooperation, and the presence of the virtualized computing platform, provides energy \mbox{self-sustainability} through the use of green energy. The \mbox{co-existence} of \mbox{on-grid} and \mbox{off-grid} BSs, in the considered scenario, provides network connectivity all the time. Towards energy allocation and routing processes, a Lyapunov-based algorithm and heuristics are used.
Numerical results, obtained with real-world energy and traffic load traces, demonstrate that the proposed algorithm (Lypunov) achieves energy transfers between $38\%$-$39\%$ and it is not affected by the control parameter impact.

\bibliographystyle{IEEEtran}
\scriptsize
\bibliography{biblio_rev}
\end{document}